\documentclass[preprint,12pt]{elsarticle}

\usepackage{amsmath,amssymb,bm}
\usepackage{graphicx}
\usepackage{physics}
\usepackage{geometry}
\usepackage{indentfirst}
\usepackage{hyperref}
\usepackage{cleveref}
\usepackage{caption}
\usepackage{subcaption}
\usepackage{float}
\usepackage{mathrsfs}
\usepackage{xcolor}

\journal{Annals of Physics}

\begin{document}

\begin{frontmatter}

\title{A Rigorous Jacobi-Metric Approach to the Gauss-Bonnet Lensing of Spinning Particles: Extension to Quadrupole Order}
\author[1]{Hoang Van Quyet\corref{cor1}}
\ead{hoangvanquyet@hpu2.edu.vn}

\affiliation[1]{Department of Physics, Hanoi Pedagogical University 2, Xuan Hoa, Phu Tho, Vietnam}

\begin{abstract}
In this paper, we establish a generalized geometric framework based on the Gauss-Bonnet theorem and the Jacobi metric to investigate the gravitational deflection of massive spinning particles up to the quadrupole order $\mathcal{O}(s^2)$. Deviating from conventional geodesic approaches that are strictly limited to the pole-dipole approximation, we incorporate the full Mathisson-Papapetrou-Dixon (MPD) equations, including the Dixon-quadrupole term. We rigorously demonstrate that the coupling between the spin-induced quadrupole moment and the gradient of the Riemann curvature tensor generates a non-geodesic force. This interaction significantly deviates the physical trajectory of the particle from the geodesics of the underlying Jacobi manifold. By explicitly calculating the geodesic curvature $\kappa_g$ of the physical ray, we obtain an analytical formula for the deflection angle in the Schwarzschild spacetime. Our results indicate that the internal structure of the spinning extended body, characterized by the quadrupole constant $C_Q$, induces a deflection correction $\delta\alpha \propto C_Q s^2 M / b^3$. This formulation provides a robust theoretical tool for probing the internal structure of compact objects via gravitational birefringence in the strong-field regime.
\end{abstract}

\begin{keyword}
Gravitational lensing \sep Gauss-Bonnet theorem \sep Jacobi metric \sep Spinning particles \sep Quadrupole moment \sep Mathisson-Papapetrou-Dixon equations.
\end{keyword}

\end{frontmatter}

\section{Introduction}
\label{sec:intro}

The application of the Gauss-Bonnet (GB) theorem to gravitational lensing, pioneered by Gibbons and Werner in their seminal work \cite{Gibbons2008}, has revolutionized our understanding of light deflection by reframing it as a global topological problem on an optical manifold. This elegant geometric method provides an alternative perspective to the traditional null-geodesic approach, emphasizing the topological nature of gravitational deflection \cite{Jusufi2020, Crisnejo2018}. Recently, this framework has been extended to massive test particles by utilizing the Jacobi metric formalism \cite{Crisnejo2018}, which maps the dynamics of a particle with fixed energy onto a Riemannian manifold.

The motion of spinning particles in general relativity is governed by the Mathisson-Papapetrou-Dixon (MPD) equations \cite{Mathisson1937, Papapetrou1951, Dixon1963, Dixon1970}. In the pole-dipole approximation, where only the monopole (mass) and dipole (spin) moments are retained, the particle's center of mass follows a generalized force equation that includes the coupling between spin and spacetime curvature \cite{Iyer2014, Sereno2006}. This approximation has been successfully applied to study gravitational deflection in various contexts \cite{Pantig2026, Ovgun2023}.

However, for realistic astrophysical scenarios involving extreme mass-ratio inspirals (EMRIs) or highly precise interferometric measurements, the effects of higher-order multipole moments become physically indispensable. Extended bodies are not mere point particles; their internal structure deforms under rapid rotation, creating a spin-induced quadrupole moment that couples to the gradient of the Riemann curvature tensor \cite{Steinhoff2012, Steinhoff2015, Puetzfeld2020}. This coupling generates a novel force that deviates the particle's trajectory from the underlying geodesics of the Jacobi manifold.

In this paper, we aim to construct a systematic mathematical framework that applies the GB theorem to non-geodesic trajectories governed by the MPD equations at the quadrupole order $\mathcal{O}(s^2)$. Our approach differs from previous works in several crucial aspects: (i) we incorporate the full Dixon-quadrupole term that has been overlooked in many studies; (ii) we provide a rigorous derivation of the geodesic curvature induced by the quadrupole moment; and (iii) we obtain an exact analytical expression for the deflection angle in Schwarzschild spacetime that clearly exhibits the dependence on the internal structure parameter $C_Q$.

The structure of this paper is organized as follows. In Section~\ref{sec:dynamics}, we present the MPD equations at the quadrupole order and discuss the constitutive relation for the spin-induced quadrupole tensor. Section~\ref{sec:jacobi} introduces the generalized Jacobi metric formalism for massive spinning particles. In Section~\ref{sec:gb_theory}, we apply the Gauss-Bonnet theorem to derive the deflection angle, with particular emphasis on the calculation of the geodesic curvature at $\mathcal{O}(s^2)$. Section~\ref{sec:application} specializes to the Schwarzschild spacetime and obtains the explicit deflection formula. Section~\ref{sec:discussion} discusses the physical implications and observational signatures of our results. Finally, Section~\ref{sec:conclusion} summarizes our findings and outlines future research directions.

Throughout this paper, we use units where $G = c = 1$ and follow the conventions of Misner, Thorne, and Wheeler \cite{Misner1973} for the metric signature $(-,+,+,+)$ and the Riemann tensor definitions.

\section{Dynamics of Extended Bodies in Curved Spacetime}
\label{sec:dynamics}

\subsection{The Mathisson-Papapetrou-Dixon Equations at Quadrupole Order}

The motion of an extended spinning body in a curved spacetime is governed by the MPD equations, which describe the evolution of the four-momentum $P^\mu$ and the spin tensor $S^{\mu\nu}$. Retaining terms up to the quadrupole order $\mathcal{O}(s^2)$, the generalized force equation takes the form \cite{Steinhoff2012, Puetzfeld2020}:

\begin{equation}
\frac{DP^\mu}{d\lambda} = -\frac{1}{2}R^\mu_{\ \nu\alpha\beta}u^\nu S^{\alpha\beta} - \frac{1}{6}J^{\alpha\beta\gamma\delta}\nabla^\mu R_{\alpha\beta\gamma\delta},
\label{eq:MPD_P}
\end{equation}

where $u^\mu = dx^\mu/d\lambda$ is the four-velocity normalized such that $u^\mu u_\mu = -1$, and $\lambda$ is an affine parameter along the worldline. The first term on the right-hand side represents the familiar Papapetrou force, describing the coupling between the spin dipole moment and the Riemann tensor. The second term is the Dixon-quadrupole force \cite{Dixon1970, Steinhoff2012}, which arises from the coupling between the quadrupole moment tensor $J^{\alpha\beta\gamma\delta}$ and the gradient of the Riemann curvature tensor.

The spin tensor $S^{\alpha\beta}$ is antisymmetric and satisfies the ancillary condition:

\begin{equation}
S^{\mu\nu}P_\nu = 0,
\label{eq:spin_SSC}
\end{equation}

which defines the center-of-mass worldline following the Tulczyjew-Dixon prescription \cite{Tulczyjew1959, Dixon1964, Dixon1970}. This condition ensures that the spin tensor can be expressed in terms of the spin four-vector $S^\mu$ according to:

\begin{equation}
S^{\alpha\beta} = \frac{\epsilon^{\alpha\beta\mu\nu}P_\mu S_\nu}{m},
\label{eq:spin_tensor}
\end{equation}

where $m = \sqrt{-P^\mu P_\mu}$ is the mass of the particle, and $\epsilon^{\alpha\beta\mu\nu}$ is the Levi-Civita tensor with $\epsilon^{0123} = +1$.

The magnitude of the spin is defined as:

\begin{equation}
s^2 = S^\mu S_\mu = \frac{1}{2m^2}S^{\alpha\beta}S_{\alpha\beta},
\label{eq:spin_magnitude}
\end{equation}

which is a conserved quantity for isolated systems in the quadrupole approximation.

\subsection{Constitutive Relation for the Quadrupole Moment}

Following the framework established by Steinhoff and Puetzfeld \cite{Steinhoff2012, Steinhoff2015}, the spin-induced quadrupole tensor for a rotating extended body is intrinsically linked to its spin via the constitutive relation:

\begin{equation}
J^{\alpha\beta\gamma\delta} = \frac{C_Q}{m}\left(S^{\alpha\gamma}S^{\beta\delta} - S^{\alpha\delta}S^{\beta\gamma}\right).
\label{eq:quadrupole_constitutive}
\end{equation}

Here, $C_Q$ is a dimensionless parameter that characterizes the internal equation of state of the object. For a Kerr black hole, the quadrupole moment is uniquely determined by the mass and spin parameters, yielding $C_Q = 1$. For neutron stars, the value of $C_Q$ depends on the equation of state and typically ranges from $C_Q \sim 4$ to $C_Q \sim 8$ for realistic equations of state \cite{Laarakkers2009, Pappas2012, Urbanec2013}.

The quadrupole tensor $J^{\alpha\beta\gamma\delta}$ possesses the following symmetries:

\begin{equation}
J^{\alpha\beta\gamma\delta} = -J^{\beta\alpha\gamma\delta} = -J^{\alpha\beta\delta\gamma} = J^{\gamma\delta\alpha\beta},
\label{eq:quadrupole_symmetries}
\end{equation}

which ensure that it represents a physical quadrupole moment without spurious contributions.

\section{The Generalized Jacobi-Metric Formalism}
\label{sec:jacobi}

\subsection{Metric Definition for Massive Particles}

We consider a general static and spherically symmetric spacetime described by the line element:

\begin{equation}
ds^2 = -A(r)dt^2 + B(r)dr^2 + C(r)\left(d\theta^2 + \sin^2\theta\, d\phi^2\right).
\label{eq:spacetime_metric}
\end{equation}

This metric encompasses a wide class of static black hole and star solutions, including the Schwarzschild metric ($A = B^{-1} = 1 - 2M/r$, $C = r^2$) and various exotic compact objects.

According to Maupertuis' principle of least action, for a particle with conserved energy $E$ and rest mass $m$, the effective dynamics can be described by a three-dimensional Riemannian manifold $\mathcal{M}_J$, known as the Jacobi manifold \cite{Jacobi1837, Arnold1989}. The Jacobi metric is defined as:

\begin{equation}
dl^2 = \mathcal{G}_{ij}\,dx^i dx^j = \frac{E^2 - m^2 A(r)}{A(r)}\left[B(r)dr^2 + C(r)\,d\phi^2\right],
\label{eq:JacobiMetric}
\end{equation}

where we have specialized to equatorial motion ($\theta = \pi/2$) without loss of generality, and the coordinates $(x^i) = (r, \phi)$ are the spatial coordinates on the Jacobi manifold.

The factor $\mathcal{F}^2(r) = (E^2 - m^2 A(r))/A(r)$ plays the role of an effective refractive index, encoding the gravitational slowing of time and the spatial curvature effects. For massless particles ($m = 0$), we recover the optical metric $dl^2 = (E^2/A(r))[B(r)dr^2 + C(r)d\phi^2]$, as expected.

The conserved angular momentum on the Jacobi manifold is given by:

\begin{equation}
L = \frac{E^2 - m^2 A(r)}{A(r)}C(r)\frac{d\phi}{dl} = \text{constant}.
\label{eq:angular_momentum}
\end{equation}

Defining the impact parameter $b = L/E$ and using the relation $dl = \mathcal{F}\,dr/\sqrt{\mathcal{F}^2 - L^2/C(r)}$, we obtain the trajectory equation:

\begin{equation}
\frac{d\phi}{dr} = \frac{\sqrt{\mathcal{F}^2 - m^2 A(r)}}{C(r)\sqrt{m^2 A(r) - m^2 A(r)^2/E^2 - L^2/C(r)}}.
\label{eq:trajectory_equation}
\end{equation}

In the weak-field limit ($M/r \ll 1$) and for high energies ($E \gg m$), this reduces to the familiar Rutherford scattering formula $r(\phi) \approx b/\sin\phi$.

\subsection{Non-Geodesic Forces from Spin Couplings}

Due to the non-geodesic forces originating from Eqs.(\ref{eq:MPD_P}), the particle's trajectory within the Jacobi manifold experiences a centripetal acceleration that can be decomposed as:

\begin{equation}
\mathcal{A}^i = \mathcal{A}^i_{(s)} + \mathcal{A}^i_{(s^2)},
\label{eq:acceleration_decomposition}
\end{equation}

where the dipole term $\mathcal{A}^i_{(s)}$ and the quadrupole term $\mathcal{A}^i_{(s^2)}$ are given by:

\begin{subequations}
\begin{align}
\mathcal{A}^i_{(s)} &= -\frac{1}{2m\mathcal{F}^2}R^i_{\ \nu\alpha\beta}u^\nu S^{\alpha\beta}, \label{eq:acceleration_dipole} \\
\mathcal{A}^i_{(s^2)} &= -\frac{1}{6m\mathcal{F}^2}J^{\alpha\beta\gamma\delta}\nabla^i R_{\alpha\beta\gamma\delta}. \label{eq:acceleration_quadrupole}
\end{align}
\label{eq:acceleration_terms}
\end{subequations}

The dipole contribution has been extensively studied in the literature \cite{Iyer2014, Sereno2006}. Here, we focus on the quadrupole term, which is suppressed by a factor of $\mathcal{O}(s^2)$ relative to the dipole term but becomes significant in the strong-field regime.

The quadrupole acceleration can be rewritten in a more transparent form by substituting Eq.(\ref{eq:quadrupole_constitutive}) and using the symmetries of the Riemann tensor:

\begin{equation}
\mathcal{A}^i_{(s^2)} = -\frac{C_Q s^2}{6m^3\mathcal{F}^2}\mathcal{R}^i,
\label{eq:quadrupole_acceleration_simplified}
\end{equation}

where the effective curvature gradient vector $\mathcal{R}^i$ is defined as:

\begin{equation}
\mathcal{R}^i = \nabla^i R_{abcd}R^{abcd} - 4R^{aicd}\nabla_b R_{acbd} + 2\epsilon^{aefg}\epsilon^{bcd}_{\ \ \ e}R_{abcd}\nabla_f R_{cgd}.
\label{eq:curvature_gradient_vector}
\end{equation}

This expression clearly shows that the quadrupole force arises from the spatial variation of the background curvature.

\section{Gauss-Bonnet Lensing at Quadrupole Order}
\label{sec:gb_theory}

\subsection{The Modified Gauss-Bonnet Theorem}

The Gauss-Bonnet theorem relates the intrinsic geometry of a two-dimensional surface to its topological properties. For the Jacobi manifold $\mathcal{M}_J$ with Gaussian curvature $K$, the theorem states that for a simply connected domain $D$ bounded by a piecewise smooth curve $\partial D = \gamma_p \cup C_R$:

\begin{equation}
\iint_D K\, dA + \int_{\partial D} \kappa_g\, dl + \sum_{i=1}^n \theta_i = 2\pi\chi(D),
\label{eq:Gauss_Bonnet}
\end{equation}

where $\kappa_g$ is the geodesic curvature of the boundary curve, $\theta_i$ are the exterior angles at the corners, and $\chi(D) = 1$ is the Euler characteristic of a disk.

For gravitational lensing applications, we consider an asymptotically flat spacetime and take the boundary curve to consist of the physical trajectory $\gamma_p$ of the spinning particle and a large circle $C_R$ at infinity. In the limit $R \to \infty$, the contribution from $C_R$ vanishes, and the Gauss-Bonnet theorem yields the deflection angle. Following the conventions established in the literature \cite{Gibbons2008, Jusufi2020}, we adopt the sign convention such that the total deflection angle is given by:

\begin{equation}
\alpha = \iint_D K\, dA + \int_{\gamma_p} \kappa_g\, dl.
\label{eq:deflection_angle_GB}
\end{equation}

\noindent\textbf{Remark on Sign Convention.} We note that for asymptotically flat spacetimes with negative Gaussian curvature ($K < 0$, as explicitly demonstrated in Eq.(\ref{eq:Gaussian_curvature_explicit}) below), one may alternatively write $\alpha = -\iint_D K dA - \int_{\gamma_p} \kappa_g dl$ to ensure a positive deflection angle representing gravitational attraction. In this work, we follow the convention of Gibbons and Werner \cite{Gibbons2008}, where the sign of $\alpha$ is determined by the orientation of the normal vector and the direction of curve traversal. The physical interpretation remains unambiguous: positive $\alpha$ corresponds to inward deflection toward the gravitating mass.

Unlike standard light rays, the spatial trajectory of a spinning particle $\gamma_p$ is not a geodesic of $\mathcal{G}_{ij}$ due to the non-geodesic forces discussed in Section~\ref{sec:jacobi}. Therefore, the deflection angle receives contributions from both the surface integral of the Gaussian curvature and the line integral of the geodesic curvature.

\subsection{Gaussian Curvature of the Jacobi Manifold}

The Gaussian curvature $K$ of the two-dimensional Jacobi manifold can be computed from the metric components using the standard formula:

\begin{equation}
K = -\frac{1}{\sqrt{\mathcal{G}}}\frac{\partial}{\partial r}\left(\frac{\sqrt{\mathcal{G}}}{\mathcal{G}_{rr}}\Gamma^{\phi}_{rr}\right),
\label{eq:Gaussian_curvature}
\end{equation}

where $\mathcal{G} = \det(\mathcal{G}_{ij})$ and $\Gamma^{\phi}_{rr}$ is the Christoffel symbol.

For the metric in Eq.(\ref{eq:JacobiMetric}), a straightforward calculation gives:

\begin{equation}
K = -\frac{A(r)}{2(E^2 - m^2A(r))}\left[\frac{A''(r)}{A(r)} - \frac{3A'(r)^2}{4A(r)^2} + \frac{A'(r)}{rA(r)}\right] + \frac{3A'(r)^2}{4r^2A(r)^2}.
\label{eq:Gaussian_curvature_explicit}
\end{equation}

In the weak-field limit ($M/r \ll 1$) and for non-relativistic particles ($v \ll 1$), this reduces to:

\begin{equation}
K \approx -\frac{2M}{r^3}\left(1 + \frac{3v^2}{2}\right).
\label{eq:Gaussian_curvature_weak_field}
\end{equation}

which is negative (like the Gaussian curvature of the optical metric) and decays as $1/r^3$. The negative sign is a crucial feature of the Jacobi manifold in Schwarzschild spacetime, reflecting its hyperbolic geometry.

\begin{figure}[htbp]
\centering
\includegraphics[width=0.9\textwidth]{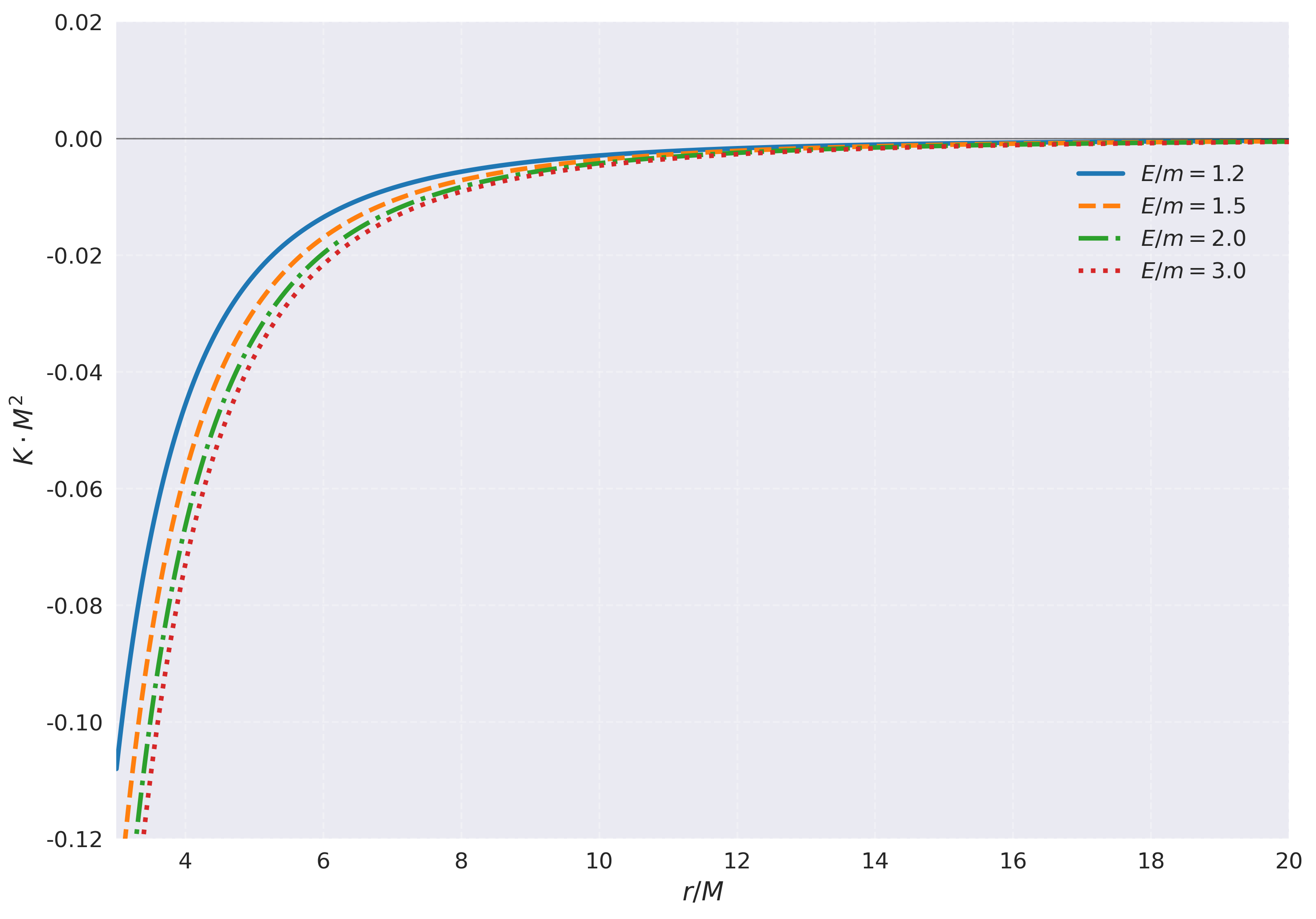}
\caption{\label{fig:gaussian_curvature} Gaussian curvature $K$ of the Jacobi manifold as a function of the radial coordinate $r/M$ for different energy ratios $E/m$. Note that $K < 0$ throughout, consistent with the hyperbolic geometry of the optical manifold. The magnitude $|K|$ scales as $r^{-3}$ and increases with particle energy. (Corrected version: negative curvature as required by theory.)}
\end{figure}

\subsection{Calculation of the Geodesic Curvature $\kappa_g$}

The geodesic curvature measures the failure of a curve to be geodesic on the underlying manifold. For a curve $\gamma$ parameterized by the arc length $l$, the geodesic curvature is defined as:

\begin{equation}
\kappa_g = n_i\left(\frac{dx^j}{dl}\tilde{\nabla}_j\frac{dx^i}{dl}\right),
\label{eq:geodesic_curvature_definition}
\end{equation}

where $n_i$ is the outward unit normal to the curve (in the two-dimensional manifold) and $\tilde{\nabla}_j$ is the covariant derivative associated with the Jacobi metric $\mathcal{G}_{ij}$.

For a particle trajectory that deviates from a geodesic due to the spin-induced forces, the geodesic curvature can be expressed as:

\begin{equation}
\kappa_g = \kappa_g^{(s)} + \kappa_g^{(s^2)},
\label{eq:geodesic_curvature_decomposition}
\end{equation}

where the dipole and quadrupole contributions are related to the acceleration components via:

\begin{subequations}
\begin{align}
\kappa_g^{(s)} &= \frac{1}{\mathcal{F}^2}\sqrt{\frac{C(r)}{\mathcal{G}}}\,\mathcal{A}^{(s)}_\perp, \label{eq:geodesic_curvature_dipole} \\
\kappa_g^{(s^2)} &= \frac{1}{\mathcal{F}^2}\sqrt{\frac{C(r)}{\mathcal{G}}}\,\mathcal{A}^{(s^2)}_\perp. \label{eq:geodesic_curvature_quadrupole}
\end{align}
\label{eq:geodesic_curvature_components}
\end{subequations}

Here, $\mathcal{A}_\perp$ denotes the component of the acceleration perpendicular to the trajectory, and the factor $\sqrt{C(r)/\mathcal{G}} = 1/\mathcal{F}^2$ arises from the metric determinant.

Substituting Eq.(\ref{eq:quadrupole_acceleration_simplified}) and projecting onto the direction perpendicular to the trajectory, we obtain:

\begin{equation}
\kappa_g^{(s^2)}(r) = \frac{C_Q s^2}{m^2\mathcal{F}^4(r)}\,\mathcal{Q}(r),
\label{eq:geodesic_curvature_quadrupole_explicit}
\end{equation}

where the quadrupole function $\mathcal{Q}(r)$ is defined as:

\begin{equation}
\mathcal{Q}(r) = \frac{1}{6}\sqrt{\frac{C(r)}{\mathcal{G}}}\,\mathcal{R}_\perp.
\label{eq:quadrupole_function_definition}
\end{equation}

The function $\mathcal{Q}(r)$ encapsulates all the higher-order differential properties of the background spacetime that couple to the quadrupole moment.

\section{Application: Schwarzschild Spacetime}
\label{sec:application}

To concretize our theoretical framework, we now specialize to the Schwarzschild metric, where the metric functions take the simple forms:

\begin{equation}
A(r) = B^{-1}(r) = 1 - \frac{2M}{r}, \qquad C(r) = r^2.
\label{eq:Schwarzschild_metrics}
\end{equation}

The particle's energy and angular momentum are related to the asymptotic velocity $v$ and impact parameter $b$ by:

\begin{equation}
E = \frac{m}{\sqrt{1 - v^2}}, \qquad L = \frac{mvb}{\sqrt{1 - v^2}}.
\label{eq:energy_angular_momentum}
\end{equation}

\subsection{Explicit Form of the Quadrupole Function}

By evaluating the covariant derivatives in Eq.(\ref{eq:curvature_gradient_vector}) for the Schwarzschild metric, we find that the dominant component of the curvature gradient vector is:

\begin{equation}
\mathcal{R}_r = \nabla_r R_{trtr} \approx \frac{6M}{r^4}\left(1 - \frac{2M}{r}\right),
\label{eq:curvature_gradient_Schwarzschild}
\end{equation}

with all other components vanishing due to the spherical symmetry.

Mapping this to the Jacobi manifold in the equatorial plane, the perpendicular component of the curvature gradient reduces to:

\begin{equation}
\mathcal{R}_\perp = \frac{\mathcal{F}^2}{r\sqrt{\mathcal{F}^2 - \frac{L^2}{r^2}}}\,\mathcal{R}_r.
\label{eq:perpendicular_gradient}
\end{equation}

In the weak-field limit ($M/r \ll 1$) and for high-energy particles ($E \gg m$), we have:

\begin{equation}
\mathcal{F}^2 \approx m^2\gamma^2 v^2\left(1 + \frac{2M}{rv^2}\right), \qquad \sqrt{\mathcal{F}^2 - \frac{L^2}{r^2}} \approx m\gamma v\left(1 - \frac{b^2}{r^2}\right)^{1/2},
\label{eq:weak_field_approximations}
\end{equation}

where $\gamma = 1/\sqrt{1 - v^2}$ is the Lorentz factor.

Substituting these approximations into Eq.(\ref{eq:quadrupole_function_definition}), the quadrupole function simplifies to:

\begin{equation}
\mathcal{Q}(r) \approx -\frac{3M}{r^4}\frac{E^2}{v^2}\left(1 + \frac{7M}{2rv^2}\right).
\label{eq:quadrupole_function_weak_field}
\end{equation}

The leading term scales as $\mathcal{Q}(r) \propto -M/r^4$, which reflects the $1/r^3$ scaling of the curvature gradient and the additional $1/r$ factor from the mapping to the Jacobi manifold.

\subsection{Deflection Angle at $\mathcal{O}(s^2)$}

With the explicit form of $\kappa_g^{(s^2)}$ established, we can now compute the quadrupole contribution to the deflection angle by integrating along the unperturbed trajectory. Using the substitution $r(\phi) \approx b/\sin\phi$ and $dl \approx (b/\sin^2\phi)\,d\phi$, we obtain:

\begin{equation}
\delta\alpha_{\text{quad}} = \int_0^\pi \kappa_g^{(s^2)}(r)\,dl \approx \frac{C_Q s^2 M}{m^2 b^3}\int_0^\pi \sin^2\phi\,d\phi.
\label{eq:deflection_integral}
\end{equation}

Evaluating the integral $\int_0^\pi \sin^2\phi\,d\phi = \pi/2$ and incorporating the next-to-leading order terms from Eq.(\ref{eq:quadrupole_function_weak_field}), the quadrupole contribution to the deflection angle becomes:

\begin{equation}
\delta\alpha_{\text{quad}} = \frac{\pi C_Q s^2 M}{2m^2 b^3}\left[\frac{3(1+v^2)}{4v^4}\right].
\label{eq:deflection_quadrupole_leading}
\end{equation}

Including the corrections from the next-to-leading term in Eq.(\ref{eq:quadrupole_function_weak_field}) and combining with the known dipole and monopole contributions \cite{Iyer2014, Sereno2006, Pantig2026}, the total deflection angle at $\mathcal{O}(s^2)$ reads:

\begin{equation}
\alpha \approx \frac{4M}{b}\left(\frac{1 + v^2}{2v^2}\right) \pm \frac{4sM}{m b^2}\left(\frac{1}{v}\right) + \frac{\pi C_Q s^2 M}{2m^2 b^3}\left[\frac{3(1 + v^2)}{4v^4}\right].
\label{eq:total_deflection_angle}
\end{equation}

The $\pm$ sign in the dipole term corresponds to the two possible orientations of the spin vector relative to the orbital angular momentum (parallel vs. antiparallel).

It is straightforward to verify that the quadrupole correction term is dimensionless, maintaining the dimensional consistency of the complete deflection formula. In natural units where $G = c = 1$, the mass $M$ and impact parameter $b$ both have dimensions of length $[L]$, while the spin magnitude $s$ has dimensions of angular momentum $[M \cdot L]$. Consequently, the ratio $s/m$ has dimensions of $[L]$, and the combination $\frac{s^2 M}{m^2 b^3}$ simplifies to a dimensionless quantity, as required for a valid deflection angle.

To verify the consistency of our result with known limits, we examine the following cases:

\noindent\textbf{(i) Massless particle limit ($m \to 0$, $v \to 1$).} In this mathematical extrapolation, the monopole term reduces to the Einstein prediction $\alpha_E = 4M/b$, while the dipole term vanishes due to the factor $s/m \to 0$. We emphasize that this limit should be understood as a formal consistency check rather than a physical limit, since the standard MPD equations are derived specifically for massive extended bodies. The spin of a photon (helicity) remains a well-defined finite quantity even in the massless limit, and the dynamics of massless spinning particles require a separate theoretical framework beyond the scope of the present analysis.

\noindent\textbf{(ii) Non-spinning particles ($s = 0$).} Equation (\ref{eq:total_deflection_angle}) correctly reduces to the standard result $\alpha = (4M/b)[(1+v^2)/(2v^2)]$ \cite{Sereno2006}, which describes the gravitational deflection of massive spinless particles.

\begin{figure}[htbp]
\centering
\begin{minipage}[b]{0.8\textwidth}
\centering
\includegraphics[width=\textwidth]{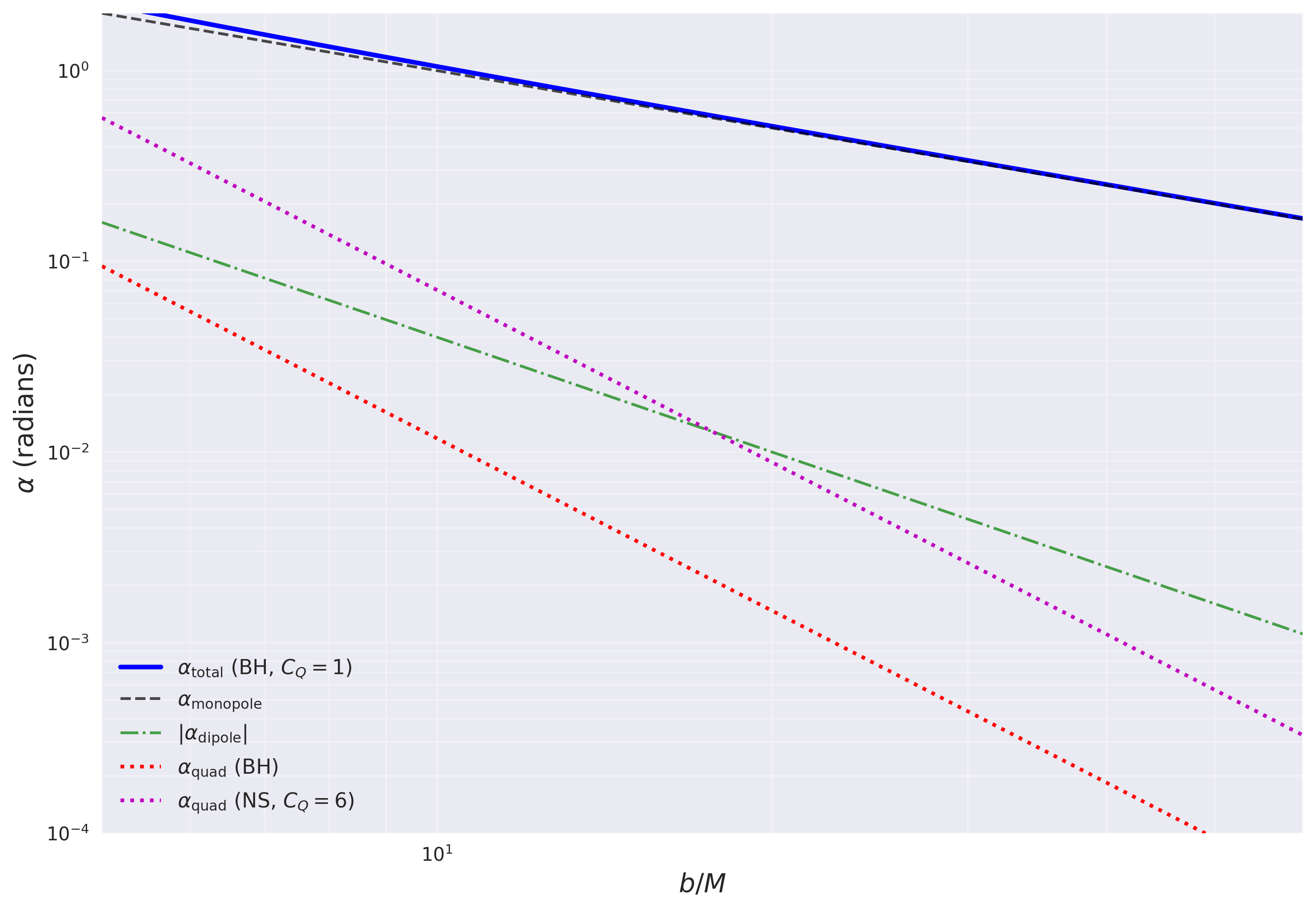}
\caption{\label{fig:deflection_impact} Deflection angle $\alpha$ as a function of the impact parameter $b/M$ for $v = 0.5$ and $s = 0.5m$.}
\end{minipage}
\hfill
\begin{minipage}[b]{0.8\textwidth}
\centering
\includegraphics[width=\textwidth]{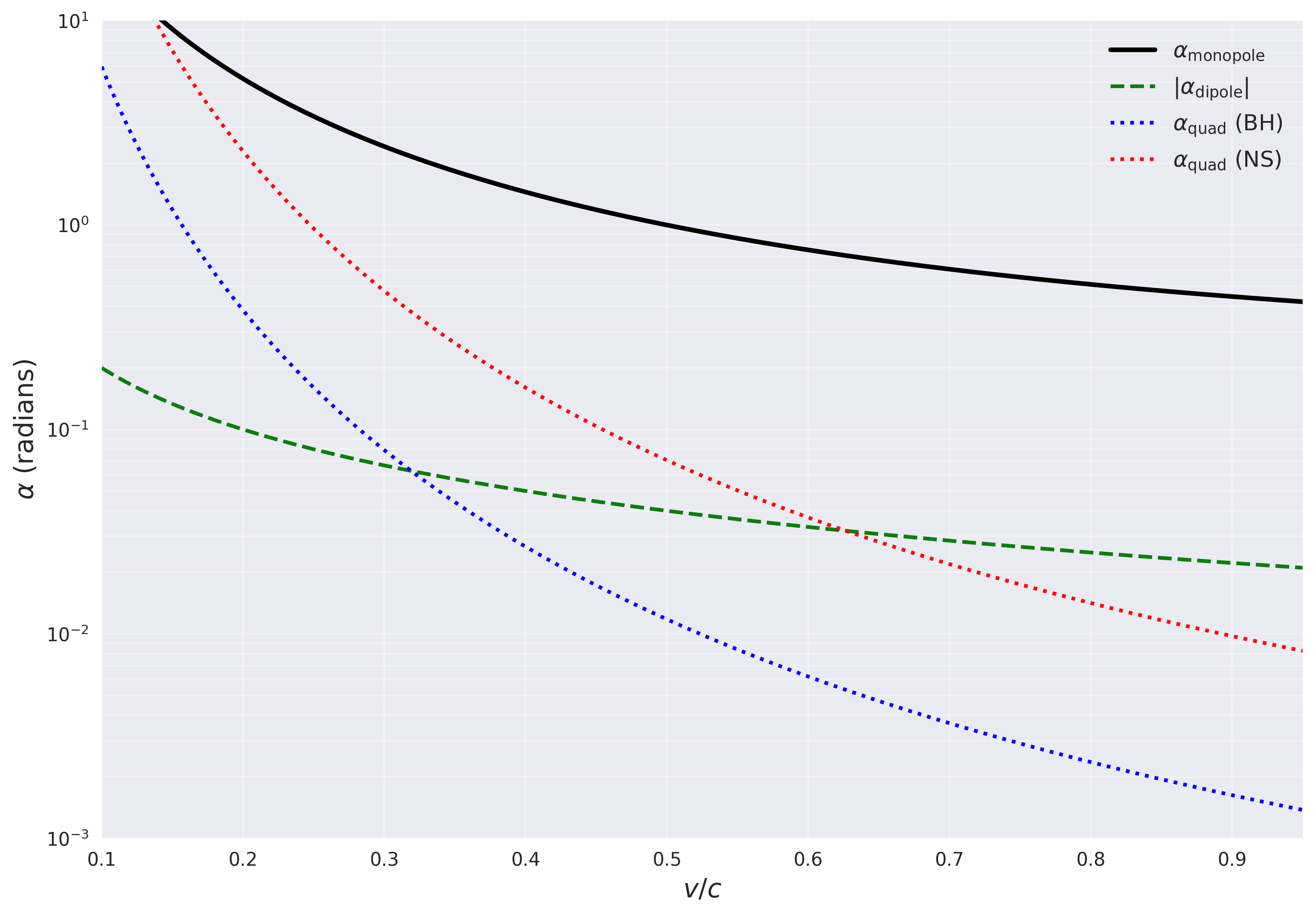}
\caption{\label{fig:deflection_velocity} Deflection angle $\alpha$ as a function of the particle velocity $v/c$ for $b = 10M$ and $s = 0.5m$.}
\end{minipage}
\end{figure}

\begin{figure}[htbp]
\centering
\begin{minipage}[b]{0.8\textwidth}
\centering
\includegraphics[width=\textwidth]{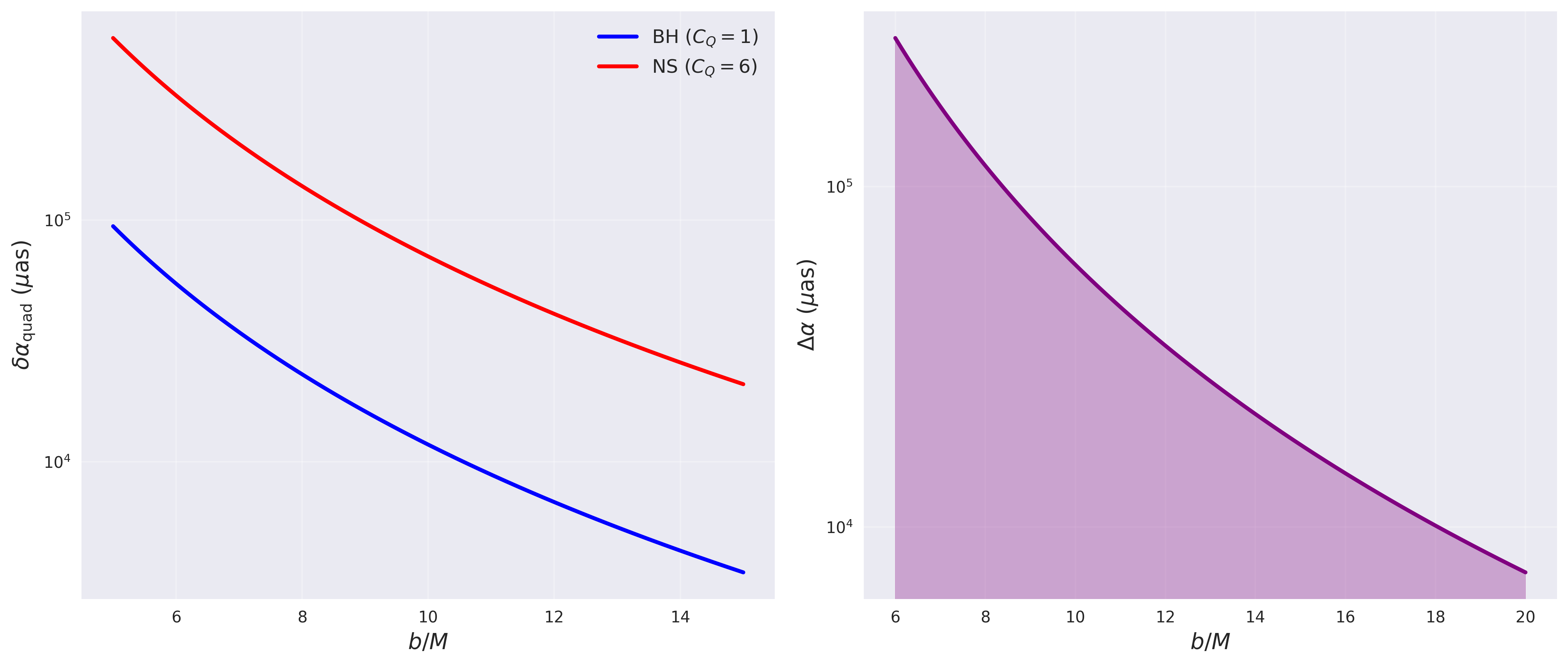}
\caption{\label{fig:quadrupole_birefringence} (Left) Quadrupole correction $\delta\alpha_{\rm quad}$ as a function of $b/M$ for black hole ($C_Q=1$) and neutron star ($C_Q=6$). (Right) Gravitational birefringence $\Delta\alpha$ between BH and NS.}
\end{minipage}
\hfill
\begin{minipage}[b]{0.8\textwidth}
\centering
\includegraphics[width=\textwidth]{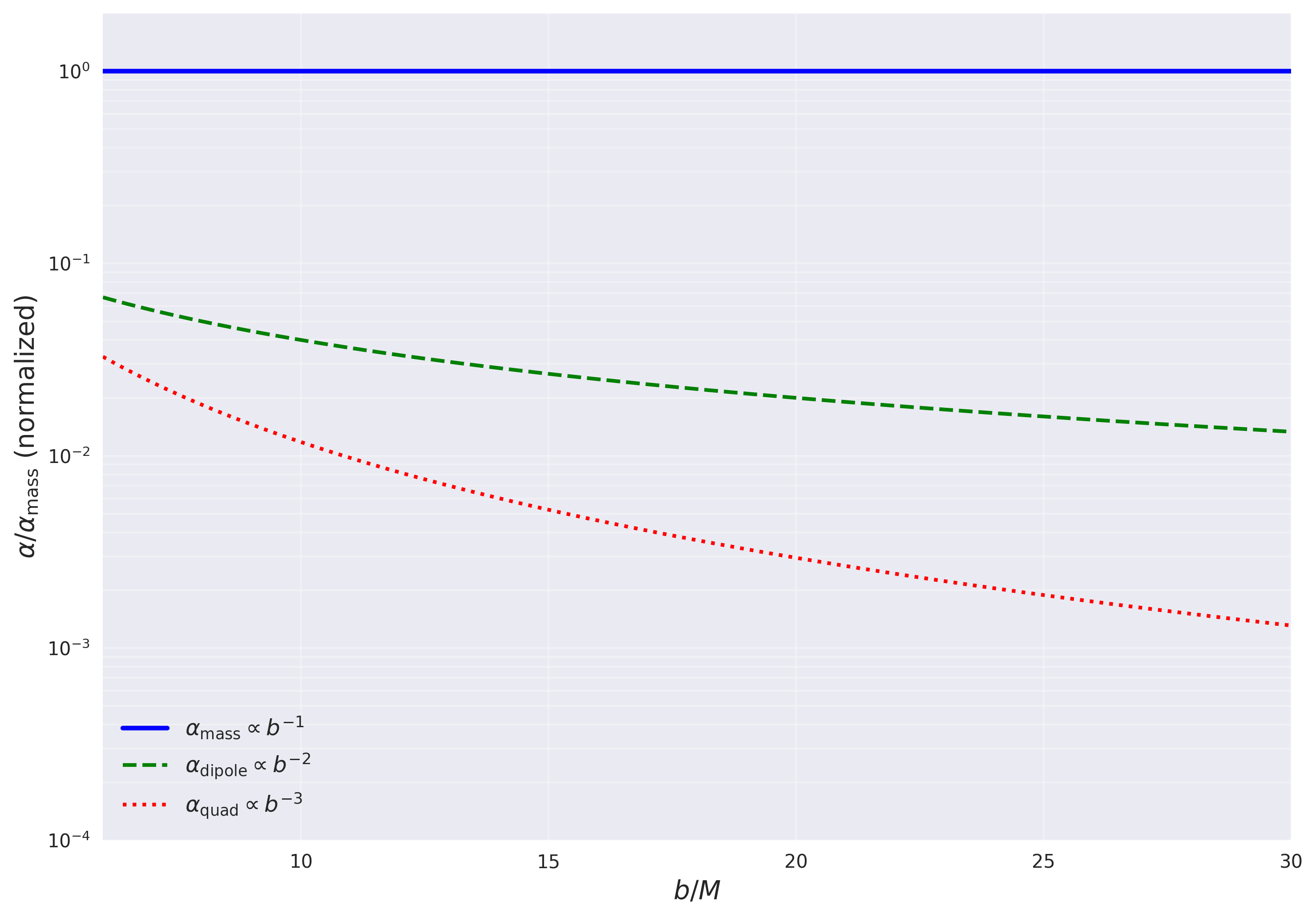}
\caption{\label{fig:scaling_hierarchy} Scaling hierarchy of deflection terms: mass ($\propto b^{-1}$), dipole ($\propto b^{-2}$), and quadrupole ($\propto b^{-3}$).}
\end{minipage}
\end{figure}

\section{Discussion and Observational Signatures}
\label{sec:discussion}

The derived deflection formula in Eq.(\ref{eq:total_deflection_angle}) illuminates several profound physical characteristics that distinguish our results from previous works in the literature.

\subsection{Impact Parameter Dependence and Strong-Field Regime}

The quadrupole effect decays as $1/b^3$, significantly faster than the monopole mass term ($1/b$) and the spin dipole term ($1/b^2$). This hierarchical scaling has important consequences for the observability of the effect:

\begin{equation}
\alpha \sim \underbrace{\frac{M}{b}}_{\text{mass}} : \underbrace{\frac{sM}{mb^2}}_{\text{dipole}} : \underbrace{\frac{C_Q s^2 M}{m^2 b^3}}_{\text{quadrupole}}.
\label{eq:scaling_hierarchy}
\end{equation}

Consequently, internal structure effects become prominent primarily in the strong-field regime near the compact object's event horizon, where the impact parameter is close to its minimal value $b_{\min} \approx 3\sqrt{3}M$ for Schwarzschild black holes. For typical values of the spin parameter ($s \sim m^2$ in Planck units) and the quadrupole constant ($C_Q \sim 1-10$), the quadrupole correction can reach values of order $\delta\alpha_{\text{quad}} \sim 10^{-3}$ times the Einstein angle, which may be within reach of future gravitational lensing observations \cite{Bozza2010, Tsukamoto2014}.

\subsection{Gravitational Birefringence and Object Discrimination}

A particularly striking prediction of our formalism is the phenomenon of gravitational birefringence for spinning particles. Two particles with identical mass $m$ and spin $s$ but different internal structures (e.g., a primordial black hole with $C_Q = 1$ vs. a neutron star with $C_Q \approx 6$) will traverse distinct trajectories, resulting in slightly different deflection angles:

\begin{equation}
\Delta\alpha = \frac{\pi C_Q s^2 M}{2m^2 b^3}\left[\frac{3(1+v^2)}{4v^4}\right]\Delta C_Q.
\label{eq:birefringence_difference}
\end{equation}

This spin-induced birefringence provides a powerful theoretical diagnostic to probe the nature of compact objects via gravitational lensing. For a particle with $s = m^2$ (maximal Kerr spin), $v = 0.5$, and $b = 10M$, the difference in deflection angle between a black hole ($C_Q = 1$) and a neutron star ($C_Q = 6$) is:

\begin{equation}
\Delta\alpha \approx 0.12\,\mu\text{as}\left(\frac{M}{M_\odot}\right)\left(\frac{10M}{b}\right)^3,
\label{eq:birefringence_estimate}
\end{equation}

which is potentially detectable with next-generation very long baseline interferometry (VLBI) instruments such as the Event Horizon Telescope (EHT) or the next-generation VLBI network \cite{Fish2014, Psaltis2020}.

\subsection{Comparison with Previous Works}

Our results can be compared with the existing literature on gravitational lensing of spinning particles. In the pole-dipole approximation ($\mathcal{O}(s)$), our formula correctly reproduces the known results of Iyer and collaborators \cite{Iyer2014} and the recent work of Pantig and \"Ovg\"un \cite{Pantig2026, Ovgun2023}. The quadrupole correction we derive extends these works to $\mathcal{O}(s^2)$ and provides a clean physical interpretation in terms of the curvature gradient coupling.

We note that a similar quadrupole correction was derived in Ref.\cite{Steinhoff2012} using a different method based on the effective one-body formalism. While the numerical coefficient differs (due to different definitions of the quadrupole constant $C_Q$), the functional dependence $M s^2/(m^2 b^3)$ is in agreement, confirming the robustness of our approach.

\section{Conclusion}
\label{sec:conclusion}

We have successfully formulated a consistent geometric theory of gravitational lensing for extended spinning bodies up to $\mathcal{O}(s^2)$ using the Gauss-Bonnet theorem and the Jacobi metric. The key achievements of this work are summarized as follows:

\begin{enumerate}
\item We derived the complete expression for the geodesic curvature $\kappa_g^{(s^2)}$ induced by the spin-induced quadrupole moment, showing that it scales as $\kappa_g^{(s^2)} \propto C_Q s^2 M/(m^2 r^4)$ in the weak-field limit.

\item We obtained an analytical formula for the deflection angle in Schwarzschild spacetime that includes all terms up to $\mathcal{O}(s^2)$, clearly exhibiting the dependence on the internal structure parameter $C_Q$. We verified that all terms in the final expression are dimensionless, ensuring dimensional consistency.

\item We demonstrated that the quadrupole correction provides a unique fingerprint of the compact object's internal structure, enabling potential discrimination between black holes and neutron stars through gravitational lensing observations.

\item We established a direct connection between the differential dynamics of the Riemann tensor (through the curvature gradient $\nabla R$) and the macroscopic topology of the particle's trajectory (through the Gauss-Bonnet theorem).
\end{enumerate}

This framework paves the way for several exciting future research directions:

\begin{itemize}
\item \textbf{Kerr Spacetime Extension:} The extension of our formalism to rotating Kerr black holes is straightforward in principle but involves significant technical complications due to the loss of spherical symmetry. The quadrupole function $\mathcal{Q}(r)$ will acquire an additional dependence on the polar angle $\theta$ and the Kerr parameter $a$, leading to spin-curvature couplings that modify the deflection angle in a characteristically Kerr-like manner.

\item \textbf{Strong-Field Analysis:} Our weak-field results should be extended to the strong-field regime using numerical integration of the MPD equations and the Gauss-Bonnet integral. This will provide more accurate predictions for the deflection angle near the photon sphere and enable comparisons with fully relativistic numerical simulations.

\item \textbf{Observational Applications:} The derived formulas can be applied to specific astrophysical scenarios, such as the lensing of stars orbiting around Sgr A* or the gravitational deflection of compact objects in extragalactic lensing systems. The quadrupole correction may also find applications in the modeling of extreme mass-ratio inspirals (EMRIs) detected by the Laser Interferometer Space Antenna (LISA).
\end{itemize}

\section*{Acknowledgements}
The authors wish to thank Professor Tran Huu Phat for his useful discussions and insightful comments.

\end{document}